\documentclass[sn-mathphys,Numbered]{sn-jnl}

\usepackage{graphicx}%
\usepackage{multirow}%
\usepackage{amsmath,amssymb,amsfonts}%
\usepackage{amsthm}%
\usepackage{mathrsfs}%
\usepackage[title]{appendix}%
\usepackage{xcolor}%
\usepackage{textcomp}%
\usepackage{manyfoot}%
\usepackage{booktabs}%
\usepackage{algorithm}%
\usepackage{algorithmicx}%
\usepackage{algpseudocode}%
\usepackage{listings}%

\begin{document}
\title{Instability in Ostwald ripening processes}

\author{\fnm{Michael}
  \sur{Wilkinson}}\email{m.wilkinson@open.ac.uk}

\affil{\orgdiv{School of Mathematics and Statistics}, \orgname{The
    Open University}, \orgaddress{\street{Walton Hall}, \city{Milton
      Keynes}, \postcode{MK7 6AA}, \country{UK}}}

\abstract{
There is a dimensionless parameter which enters 
into the equation for the evolution of supersaturation 
in Ostwald ripening processes. This parameter is typically 
a large number. Here it is argued that the consequent stiffness of 
the equation results in the evolution of the supersaturation being unstable. 
The instability is evident in numerical simulations of Ostwald ripening.
}

\maketitle

\section{Introduction}
\label{sec: 1}

Ostwald ripening \cite{Ost96} is a coarsening process which occurs after phase separation. 
A remarkable analysis by Lifshitz and Slyozov \cite{Lif+58,Lif+61} 
(see also closely related work by Wagner, \cite{Wag61}) is 
the basis of most theoretical discussions, and its predictions 
are in quite good, although not perfect, agreement with experimental observations \cite{Voo84}.
This paper will present evidence that there is an instability in the 
long-time limit of Ostwald ripening, which implies that the asymptotic 
long-time evolution is yet to be fully understood, and which suggests 
a possible explanation for the discrepancies. 

Ostwald ripening processes 
can occur in solid, liquid or gas phases. The discussion in this paper 
will assume a dilute suspension of liquid droplets in a gas phase. 
This is an idealisation of the atmospheric aerosol, which consists of microscopic water droplets 
in air. Considering this case removes many of the complications which arise when 
volume fractions of the phases are comparable.

The coarsening system is described by the distribution of 
droplet sizes, $a(t)$. At any given time there is a critical droplet size $a_{\rm cr}(t)$, 
such that droplets larger than $a_{\rm cr}(t)$ grow, and smaller droplets shrink, as 
a consequence of their higher Laplace pressure. 
It is natural to use dimensionless variables 
\begin{equation}
\label{eq: 1.0}
x(t)=\frac{a_{\rm cr}(t)}{a_{\rm cr}(0)} 
\ ,\ \ \ 
y(t)=\frac{a(t)}{a_{\rm cr}(t)}
\ .
\end{equation}
The argument which is developed in \cite{Lif+58,Lif+61} is structured in an
unusual way. They give an equation of motion for the dimensionless droplet sizes $y(t)$, but
there is no explicit equation of motion for the dimensionless typical droplet size $x(t)$.
The evolution of $x(t)$ enters into the equation of motion for $y(t)$ via a parameter $\nu$, defined by
\begin{equation}
\label{eq: 1.1}
\nu=\frac{1}{x^2\dot x}
\quad \quad \quad \quad
{\rm growth}-{\rm rate\ parameter}
\end{equation}
where $\dot x$ is the time-derivative of $x$.
It is argued \cite{Lif+58,Lif+61} that a satisfactory solution can only be obtained
if $\nu(t)\to {\rm const.}$. Moreover, it is argued that $\nu$ must approach a value
which implies that there is a \lq bottleneck' slowing the decrease of the dimensionless 
droplet sizes, implying that $\nu\to 27/4$ (the equation of motion for $y$ will be
discussed in section \ref{sec: 2}, equation (\ref{eq: 2.8})). 
Their solution predicts a distribution of droplet sizes $y$ at large time,
which is argued to be insensitive to the initial droplet size distribution.

The logic of the argument is reminiscent of a statement which occurs 
in the Sherlock Holmes stories: ``When you have eliminated all which is 
impossible, then whatever remains, however improbable, must be the truth" \cite{Con26}.
It would be reassuring to see that no other possibilities had been missed
which might also be viable as solutions. In particular, the arguments 
in \cite{Lif+58} do not address the possibility that $\nu(t)$ might 
fluctuate on a short timescale, possibly erratically. 

In the following, an equation of motion for 
$x(t)$ will be obtained, depending upon a dimensionless parameter, $\alpha$,
previously introduced in \cite{Wil25}.
In the case of the atmospheric aerosol, $\alpha\gg 1$, and there 
are arguments which suggest that this applies quite generally.  As a consequence, 
the equation of motion for $x(t)$ is \lq stiff', in the sense that small deviations of the trajectory 
will result in large corrections to $x(t)$. This stiffness could imply that the equation of motion 
is ill-conditioned for numerical solution, or even that it is fundamentally unstable. 
In section \ref{sec: 2}, it is argued that the expectation value of $y(t)$ should satisfy 
$\langle y\rangle\to 1$ as $\alpha \to \infty$ (where $\langle\cdot\rangle$ 
denotes the expectation value throughout). It is also argued that fluctuations of 
$\langle y-1\rangle$ cause fluctuations of $\nu$ which are amplified by a factor which 
is proportional to $\alpha$. 

Most treatments of Ostwald ripening, including the seminal work by Lifshitz and Slyozov \cite{Lif+58}, 
represent the population of droplets by a probability
distribution function which represents their sizes. This appears to be a reasonable approach
while the number of droplets, $N$, remains large. There will, however, be fluctuations of the 
mean relative droplet radius $y$ which are of order $1/\sqrt{N}$, associated with randomness 
of the initial droplet sizes. Because fluctuations of $\langle y-1\rangle$ cause fluctuations of $\nu$ which are 
amplified by a factor proportional to $\alpha\gg 1$, there may be large fluctuations of $\nu$, which become 
more significant as the number of droplets decreases. In section \ref{sec: 2} it is argued that there are 
large fluctuations of $\nu$ when a dimensionless variable
\begin{equation}
\label{eq: 1.2}
\Omega\equiv\frac{\alpha x}{\sqrt{N}}
\end{equation}
becomes large. This indicates that, if $\alpha\gg 1$, the Lifshitz-Slyozov 
theory always fails in the long-time limit as the number of surviving droplets decreases, and that this failure 
will be more pronounced as $\alpha$ is increased.

Because of the complexity of the equations describing Ostwald ripening, this paper 
will emphasise numerical investigations.  
The equations of motion for Ostwald ripening will be 
investigated numerically in section \ref{sec: 3}, which describes simulations 
of the Ostwald ripening process with initial sizes drawn at random from a specified 
distribution. It is demonstrated that there are erratic fluctuations of $\nu(t)$, which are
related to the counting statistics fluctuations of the droplet sizes. The droplet 
growth rate parameter $\nu$ and the radius distribution are not observed to converge to the Lifshitz-Slyozov theory,
even at very large time, although the growth of the droplet radii does follow the Lifshitz-Slyozov solution closely. 
The fluctuations of $\nu(t)$ become more severe as $\alpha$ is increased. 

The fluctuations which are evident
in these simulations is associated with modelling the droplets as discrete objects, rather than as a continuous distribution. 
This approach is relevant to experimental investigations, because the supersaturation field which the droplets 
experience is sensitive to conditions in a finite volume of fluid, which is determined by the volume of the container,
or the volume which can be explored by diffusion over the timescale of the experiment (whichever is the smaller).
Earlier numerical investigations of Ostwald ripening, see, for example \cite{Che+93,Yao+93,Car+04}, all 
appear to have treated the droplet size distribution as a continuous function, and they have not shown 
examples of the type of instability which was found in this study.

Section \ref{sec: 4} gives an equation for $\nu (t)$ which is valid as $\alpha \to \infty$. Numerical simulations show that 
the system is highly unstable in this limit, in accord with the observations in section \ref{sec: 3}.
Section \ref{sec: 5} is a summary and conclusion.

Much of the literature of Ostwald ripening assumes from the beginning that the amount of material 
in the solute phase is negligible (see, for example \cite{Pen97,Nie+99}), consistent with taking the 
$\alpha\to \infty$ limit, as well as assuming that the droplet size distribution is 
represented by a continuous function, equivalent to taking the $N\to \infty$ limit.
As these limits are approached, the variable $\Omega$ defined by (\ref{eq: 1.2}) is indeterminate. The physics 
of the problem indicates that $\alpha$ should be treated as a large parameter, and that $N$ decreases toward zero as a function 
of time, so that $\Omega$ does eventually become large, implying a breakdown of the Lifshitz-Slyozov solution.

A brief survey of significant papers on Ostwald ripening which are not related to the present study should be included here. The 
Lifshitz-Slyozov solution satisfies a similarity property \cite{Lif+58,Lif+61}, and 
analogous similarity solutions have been obtained for many other coarsening processes \cite{Rat+85}. 
More similarity solutions have been obtained for variants of the Ostwald 
ripening process which describe systems where the volume fractions of different
phases are comparable: see \cite{Bal02} and references therein. Also, various authors have considered 
the properties of a one-parameter family of similarity solutions for which $\nu\ne 27/4$ \cite{Bro89,Hil+92}.
There is a substantial literature on whether there is a selection mechanism for these solutions: for example  
\cite{Mee+96,Gir+98,Mee99,Nie+01,Nie+06}. These latter papers used reduced equations of motion, 
in which supersaturation is not a dynamical variable, following the approach of \cite{Pen97,Nie+99}.
See also \cite{Car+04} for a discussion of numerical methods appropriate for treating Ostwald ripening with a continuous 
distribution of droplet sizes, and \cite{Col+02} for a discussion of the long-time limit in that case.
The results of this work do not challenge whether these similarity solutions are mathematically correct, but they do call 
into question whether they are approached in the long-time limit of physical processes.

\section{Theoretical considerations}
\label{sec: 2}

\subsection{Fundamental equations}
\label{sec: 2.1}

The equations will be discussed in terms of the atmospheric aerosol, 
which consists of very small water droplets uniformly and randomly
dispersed in air \cite{Mas71}. The complications which arise from a finite volume 
fraction \cite{Mar+84} do not arise in this case, and effects of gravity \cite{Rat+85} will also 
be neglected.

The effects of surface tension are determined by a length scale $\Lambda$,
which depends upon the surface tension $\gamma$, the molecular
volume of water $v_{\rm m}$, and the equilibrium volume 
fraction of water molecules $\Phi_{\rm e}$
\begin{equation}
\label{eq: 2.1}
\Lambda=\frac{2\gamma v_{\rm m}}{kT}\Phi_{\rm e}
\ .
\end{equation}
For water droplets in air, 
$\Lambda\approx 2.1\times 10^{-14}{\rm m}$.
(The physical parameters used were as follows: surface tension: $\sigma=7.0\times 10^{-2}\, {\rm N\,m}^{-1}$, 
molar volume: $V_{\rm m}=10^{-3}/18\,{\rm m}^3$, implying molecular volume:  
$v_{\rm m}=9.93\times 10^{-29}\,{\rm m}^3$. Saturated air at $15\,^\circ {\rm C}$ contains 
approximately $6.0\,{\rm g\,m}^{-3}$ water vapour, corresponding to a volume fraction at equilibrium 
$\Phi_{\rm e}=6.0\times 10^{-6}$.) 

A droplet of radius $a$ can either grow or shrink depending 
upon whether the supersaturation of the gas surrounding it is greater than or less than 
$\Phi_{\rm cr}=\Lambda/a$. The equation of motion for the radius of a droplet is \cite{Gre56,Lif+61}
\begin{equation}
\label{eq: 2.2}
\frac{{\rm d}a}{{\rm d}t}=\frac{D\Lambda}{a}\left[\frac{1}{a_{\rm cr}(t)}-\frac{1}{a}\right]
\end{equation}
where the critical radius is 
\begin{equation}
\label{eq: 2.3}
a_{\rm cr}(t)=\Lambda/\Phi_{\rm s}(t)
\end{equation}
and where $\Phi_{\rm s}(t)$ is the supersaturation volume fraction, $D$ is a diffusion 
coefficient. The diffusion coefficient of water molecules in air is approximately 
$D=2.0\times 10^{-5}\,{\rm m}^2{\rm s}^{-1}$. Using this figure neglects effects of cooling 
of an evaporating droplet, requiring replacement of latent heat. A smaller effective diffusion 
coefficient $D_{\rm eff}$ should be used \cite{Mas71}, but this correction will not be applied here.

Equation (\ref{eq: 2.2}) is, in principle, to be solved for each of the initial droplets, until the 
point where $a$ reaches zero (indicating that the droplet has evaporated).

At any time, the supersaturation is
\begin{equation}
\label{eq: 2.4}
\Phi_{\rm s}=\Phi_0-\frac{4\pi}{3}\frac{N(t)}{V}\langle a^3\rangle
\end{equation}
where $N(t)$ is the number of droplets at time $t$ in a volume $V$ and $\Phi_0$ is the volume fraction of liquid at large 
time, when the supersaturation has decreased to zero. (The notation $\langle X\rangle$
will be used throughout for the expectation value of $X$ for those droplets which still
exist). As well as (\ref{eq: 2.2}), we should also consider the equation of motion for the critical radius:
\begin{equation}
\label{eq: 2.5}
\frac{{\rm d}a_{\rm cr}}{{\rm d}t}=-\frac{a_{\rm cr}^2}{\Lambda}\frac{{\rm d}\Phi_{\rm s}}{{\rm d}t}
\ .
\end{equation}
Using equations (\ref{eq: 2.2}), (\ref{eq: 2.4}):
\begin{equation}
\label{eq: 2.6}
\frac{{\rm d}\Phi_{\rm s}}{{\rm d}t}=-\frac{4\pi}{V}\sum_i a_i^2 \frac{{\rm d}a_i}{{\rm d}t}
=\frac{4\pi D\Lambda}{V}\sum_i \left(1-\frac{a_i}{a_{\rm cr}}\right)
\end{equation}
where $V$ is the volume of the system. 

\subsection{Dimensionless equations of motion}
\label{sec: 2.2}

In addition to the dimensionless variables $x$, $y$ mentioned in section \ref{sec: 1}, 
introduce a dimensionless time variable, $\tilde t$:
\begin{equation}
\label{eq: 2.7}
x(t)=\frac{a_{\rm cr}(t)}{a_0},\quad y(t)=\frac{a(t)}{a_{\rm cr}(t)}, \quad 
\tilde t=\frac{D\Lambda t}{a_0^3}
\end{equation}
where $a_0=a_{\rm cr}(0)$.

These lead to an equation of motion for the dimensionless relative 
droplet size $y(\tilde t)$
\begin{equation}
\label{eq: 2.8}
\frac{{\rm d}y}{{\rm d}\tilde t}=\frac{1}{x^3}\left[\frac{y-1}{y^2}-\tilde \nu y\right]
\end{equation}
where 
\begin{equation}
\label{eq: 2.85}
\tilde \nu\equiv x^2\frac{{\rm d}x}{{\rm d}\tilde t}=\frac{1}{\nu}
\ .
\end{equation}
Consider the form of equation (\ref{eq: 2.6}) in dimensionless 
coordinates. Defining another dimensionless constant
\begin{equation}
\label{eq: 2.9}
\alpha=\frac{4\pi N_0 a_0^4}{\Lambda V}
\end{equation}
(where $N_0=N(0)$), equations (\ref{eq: 2.5}) and (\ref{eq: 2.6}) imply that
\begin{equation}
\label{eq: 2.10}
\frac{{\rm d}x}{{\rm d}\tilde t}=\alpha x^2 \frac{1}{N_0V}\sum_i \left(y_i-1\right)
\ .
\end{equation}
Denoting the probability that a droplet survives until time $\tilde t$ by 
$P_{\rm s}(\tilde t)=N(\tilde t)/N_0$, 
the equation of motion for the dimensionless typical droplet size is 
\begin{equation}
\label{eq: 2.11}
\frac{{\rm d}x}{{\rm d}\tilde t}=\alpha P_{\rm s}(\tilde t)x^2\langle y-1\rangle
\ .
\end{equation}
Now estimate the value of $\alpha$ for the atmospheric aerosol system. 
Assume that $a_0$ takes a typical value for cloud droplets, $a_0=10^{-5}\,{\rm m}$,
and that the liquid water content is $5\%$ of the total water content \cite{Mas71}, so that 
$\Phi_0=3\times 10^{-7}$. Writing $\Phi_0=4\pi N_0a_0^3/3$ leads to 
$N_0=7 \times 10^7\,{\rm m}^{-3}$, and hence $\alpha \approx 420$.

The parameter $\alpha$ in equation (\ref{eq: 2.11}) has been shown to be large 
for the atmospheric aerosol, and large values will also obtain in other systems 
where Ostwald ripening might occur. 
If ${\rm d}x/{\rm d}\tilde t$ is a well-behaved function of the dimensionless time $\tilde t$, 
then in the limit as $\alpha\to \infty$ the distribution of values of $y$ is constrained:
\begin{equation}
\label{eq: 2.12}
\lim_{\alpha \to \infty}\langle y\rangle=1
\ .
\end{equation}

Equation (\ref{eq: 2.8}) indicates that it is advantageous to make a further change of dimensionless 
time variable: define
\begin{equation}
\label{eq: 2.13}
\frac{{\rm d}\tau}{{\rm d}\tilde t}=\frac{1}{x^3}
\ .
\end{equation}
With this new time variable, the equations of motion become
\begin{equation}
\label{eq: 2.14}
\frac{{\rm d}y}{{\rm d}\tau}=\frac{y-1}{y^2}-\tilde \nu y
\end{equation}
with
\begin{equation}
\label{eq: 2.15}
\tilde \nu=\frac{{\rm d}\ln x}{{\rm d}\tau}
=\alpha x^4 P_{\rm s}\langle y-1\rangle
\ .
\end{equation}
In the long-time limit, the supersaturation is expected to approach zero, so that 
the volume fraction of the liquid state approaches a constant, $\Phi_0$:
\begin{equation}
\label{eq: 2.17}
\Phi_0=\frac{4\pi}{3}\frac{N_0}{V} a_0^3\langle y^3\rangle P_{\rm s}x^3
\ .
\end{equation}
This expression motivates the introduction of another dimensionless quantity, 
which characterises the initial condition of the ripening process:
\begin{equation}
\label{eq: 2.18}
\beta\equiv \frac{3\Phi_0 V}{4\pi N_0a_0^3}
\ .
\end{equation}
This is number is approximately unity if most of the dissolved component has already 
condensed to form droplets at $t=0$. We have, when $\tau\to \infty$,
\begin{equation}
\label{eq: 2.19}
\tilde \nu=\alpha\beta x\frac{\langle y-1\rangle}{\langle y^3\rangle}
\ .
\end{equation}
In the following, it will be assumed that $\beta\approx 1$ and that 
$\langle y^3\rangle $ is slowly varying as a function of $\tau$, and close to unity.

Equation (\ref{eq: 2.19}) is the principal result which will be used to explain the instability
of the Ostwald ripening process. 

\subsection{Role of counting fluctuations}
\label{sec: 2.3}

Arguments in \cite{Lif+58} suggest that $\tilde \nu$ takes a value which ensures
that there is a bottleneck in the growth of droplets. These are persuasive, but not entirely compelling.
Consider how they might be revised in the light of equation (\ref{eq: 2.19}), which relates 
the growth-rate parameter $\tilde \nu$ to the distribution of the normalised droplet size, $y$.
The bottleneck, determined by finding a value of $\tilde \nu$ such that the maximum of 
${\rm d}y/{\rm d}\tau$ is equal to $0$, occurs at $\tilde \nu=4/27$. Then (\ref{eq: 2.19}) indicates that, when
$\alpha \gg 1$, $\langle y-1\rangle$ approaches a small value, inversely proportional to $\alpha$.

However, $\langle y\rangle$ is the result of averaging over a finite number of values:
\begin{equation}
\label{eq: 2.20}
\langle y\rangle=\frac{1}{N(\tau)}\sum_i y_i
\end{equation}
with the values of $y_i(\tau)$ determined from the initial droplet sizes, by 
integrating (\ref{eq: 2.14}) to time $\tau$. 
If the initial droplet sizes are determined by independent samples of a probability 
density function, there will be random fluctuations of $\langle y\rangle$.  

To estimate these fluctuations, consider the variance of (\ref{eq: 2.20}): its second moment is
\begin{equation}
\label{eq: 2.21}
\langle\langle y\rangle^2\rangle=\frac{1}{N^2}\sum_i\sum_j\langle y_iy_j\rangle=\frac{1}{N^2}N(N-1)+\frac{1}{N}\langle y^2\rangle
\end{equation}
so that the variance of $y-1$ is
\begin{equation}
\label{eq: 2.22}
{\rm Var}(y-1)=\frac{1}{N}[\langle y^2\rangle-1]
\ .
\end{equation}
This calculation indicates that there will be fluctuations of $\langle y\rangle$ which are of 
order $1/\sqrt{N}$, due to random variations of the initial droplet sizes. 

There will also be fluctuations of $\langle y\rangle$ which arise from the evaporation 
of individual droplets: every time one of the $N$ droplets evaporates, the mean value increases by 
a factor of $N/(N-1)$. This source of fluctuations makes a contribution to fluctuations 
of $\langle y\rangle$ which is of order $1/N$: smaller in the limit as $N\to \infty$, but potentially 
significant at later stages of the process.

The conclusion is that there will be random fluctuations of $\tilde \nu$ which become significant when 
the dimensionless variable defined in (\ref{eq: 1.2}), $\Omega=\alpha x/\sqrt{N}$, is large. 
For any Ostwald ripening process, this condition will be satisfied at sufficiently
large time.

\subsection{Further argument for instability}
\label{sec: 2.4}

The arguments presented above indicate that there will be fluctuations of $\tilde \nu$ 
which are a consequence of counting-number fluctuations being amplified due to the stiffness
of the equation determining the supersaturation, expressed in dimensionless form by (\ref{eq: 2.19}).
There may also be other sources of instability.

The solution suggested by \cite{Lif+58,Lif+61} depends upon the decrease of $y(t)$ being 
slowed by tuning $\nu(t)$ to cause a  \lq bottleneck', meaning a point at which 
the velocity of $y(t)$ approaches zero. The flux of $y(t)$ values at this 
bottleneck is exquisitely sensitive to the value of $\nu(t)$. However, the value of $\langle y\rangle$ is 
sensitive to the values of $y(t)$ which passed through the bottleneck 
some time ago. So there is a feedback loop which appears to have high-gain and a delayed action. 
This combination suggests that the equation for $\nu(t)$ is fundamentally unstable.
In principal, this argument can be made quantitative by writing an equation for the response  
of the mean value to small changes of the growth-rate parameter. If the distribution of $y$ has reached a 
stationary state, for small fluctuations, there is a linear 
relationship expressed via a response kernel $K(\cdot)$:
\begin{equation}
\label{eq: 2.23}
\delta \langle y\rangle (t)=\int_{-\infty}^t{\rm d}t'\ K(t-t')\,\delta \nu(t')
\ .
\end{equation}
This approach was adopted in \cite{Wil25}, which considered the stability of a modified Ostwald ripening 
model which has a steady-state solution, obtaining a stability condition expressed in terms of the 
Laplace transform of $K$. However, there are several difficulties which arise when attempting 
to extend this approach to the present problem. The most fundamental of these is the fact that, 
for the problem treated in this paper, there is no clearly-defined reference trajectory from which 
the deviations can be measured. The definition of the stability criterion via a linear response kernel 
will not be pursued in this work 

\section{Numerical studies: finite $\alpha$, random initial radii}
\label{sec: 3}

The mean-field Ostwald ripening process was investigated by a direct simulation. The simulation 
uses $N(t)$ droplets, each with dimensionless radius $a_i(t)$.  
The $a_i$ values were drawn from a specified initial distribution, with 
probability density function $p_0(a)$. The equations of motion for the radii $a_i$  and the dimensionless 
supersaturation $s$ are
\begin{equation}
\label{eq: 3.1}
\frac{{\rm d}a_i}{{\rm d}t}=\frac{s}{a_i}-\frac{1}{a_i^2}
\ ,\ \ \
\frac{{\rm d}s}{{\rm d}t}=\frac{\alpha}{N_0}\sum_i \left(1-sa_i\right)
\  .
\end{equation}
(These are dimensionless versions of equations (\ref{eq: 2.2}) and (\ref{eq: 2.6}), respectively).
The parameter $\nu$ was estimated by evaluating
\begin{equation}
\label{eq: 3.2}
\nu=-\frac{s^4}{{\rm d}s/{\rm d}t}
\end{equation}
where the time derivative of $s$ was estimated numerically. 

\begin{figure}
\begin{center}
\includegraphics[width=6.3cm]{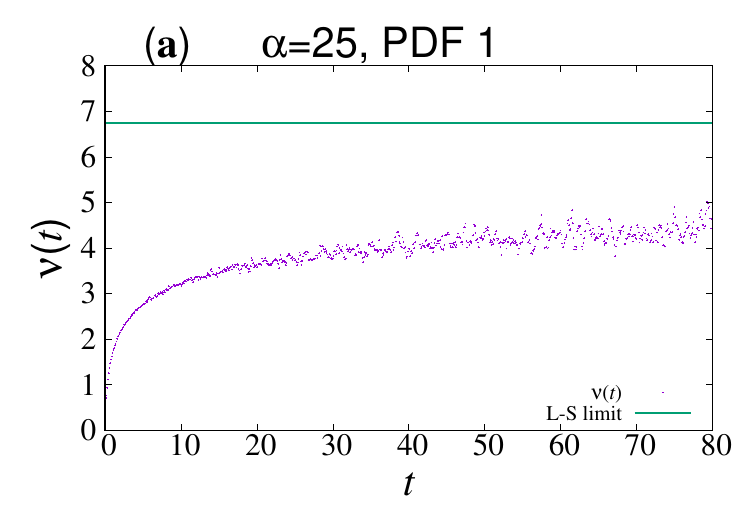}
\includegraphics[width=6.3cm]{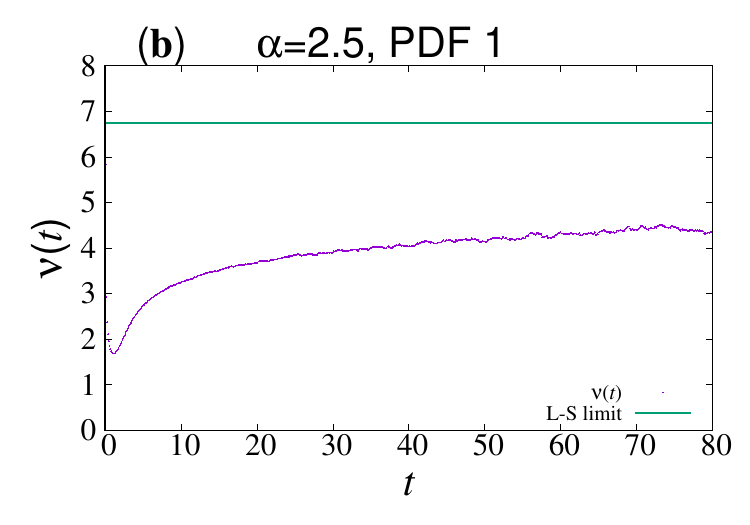}
\includegraphics[width=6.3cm]{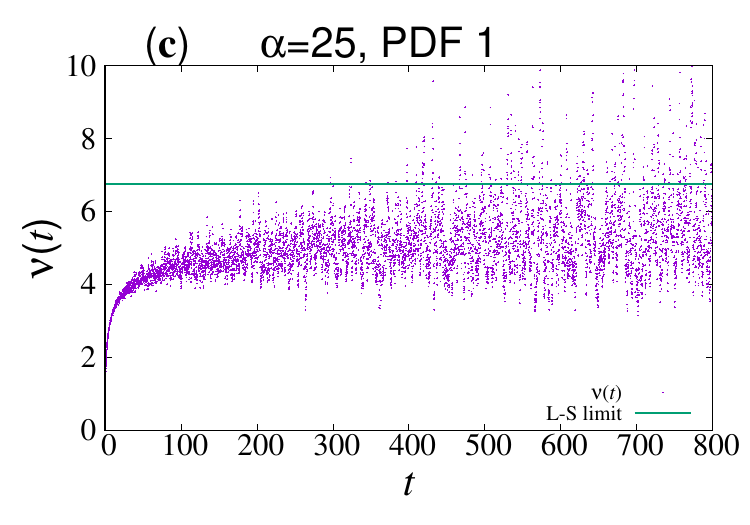}
\includegraphics[width=6.3cm]{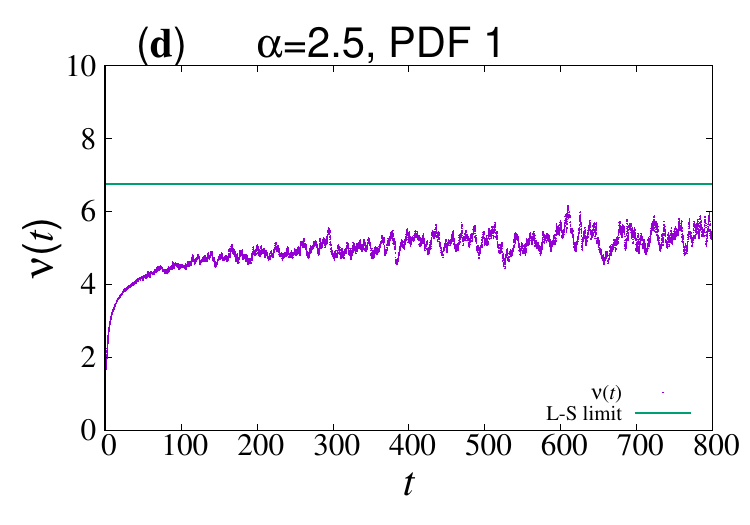}
\includegraphics[width=6.3cm]{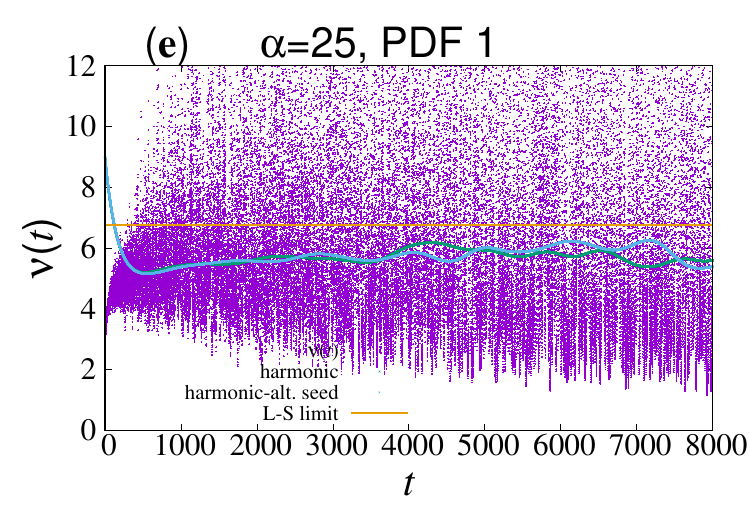}
\includegraphics[width=6.3cm]{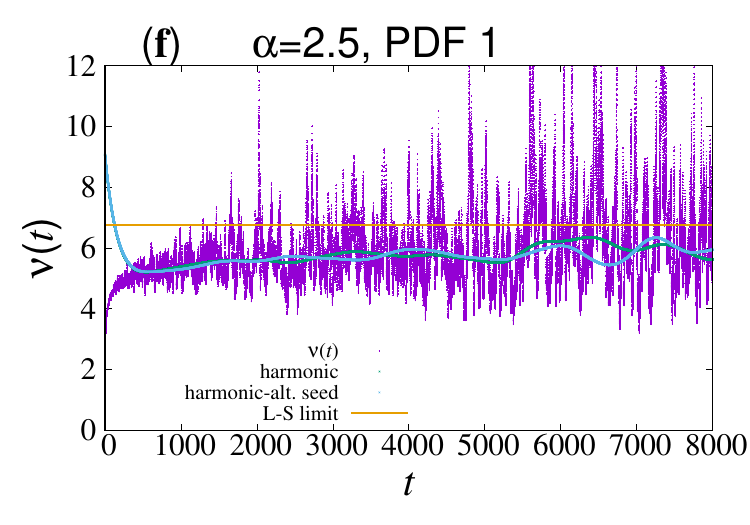}
\includegraphics[width=6.3cm]{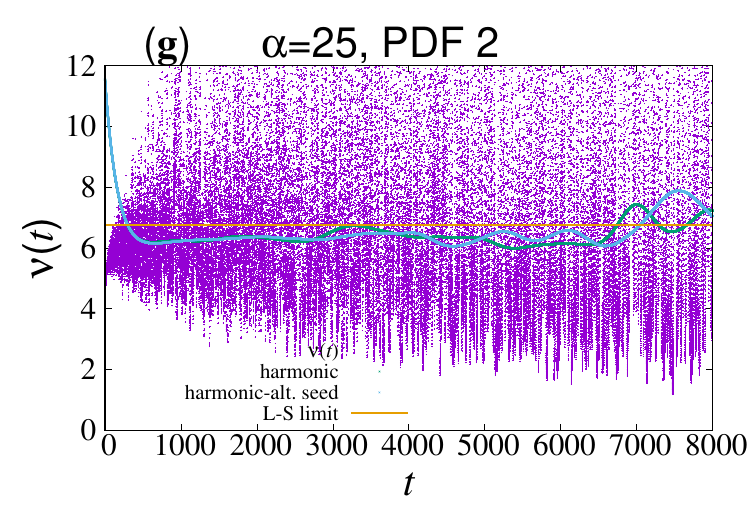}
\includegraphics[width=6.3cm]{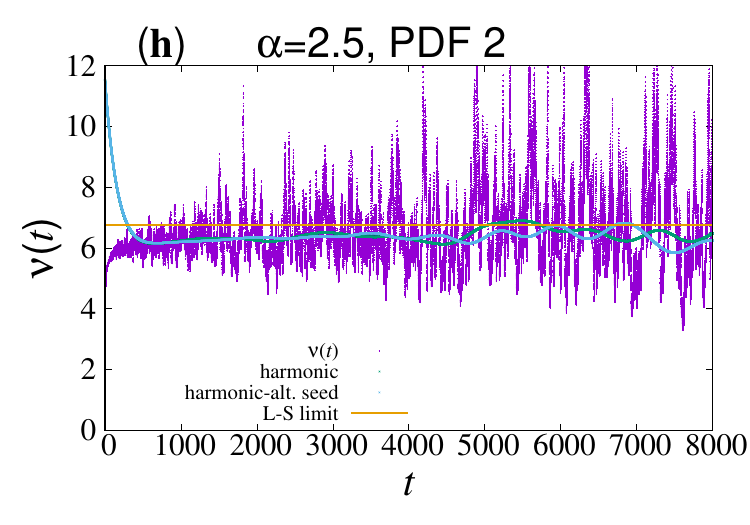}
\end{center}
\caption{
\label{fig: 1}
Numerical evaluation of $\nu(t)$, for different initial distributions $p_0(a)$ and different values of $\alpha$.
In all cases there are initially $N=10^7$ droplets. The long-time simulations also show the running harmonic mean 
$\mu(t)$ (green), including data for a different random-number seed (blue). The variance of the running weight 
was $\Delta t=250$ (see equation (\ref{eq: 3.3})). The values of $\alpha$ and the choice 
of initial radius PDF are indicated above each plot: 
{\tt PDF\ 1}, equation (\ref{eq: 3.4}) has a distribution of initial \emph{radii} with an exponential tail, 
{\tt PDF\ 2}, equation (\ref{eq: 3.5}) has a distribution of initial \emph{volumes} with an exponential tail.
}
\end{figure}

Equation (\ref{eq: 1.1}) implies that 
\begin{equation}
\label{eq: 3.25}
x^3(t)=x^3(0)+3\int_0^t{\rm d}t'\ \frac{1}{\nu(t')}
\end{equation}
so that, if the value of $\nu(t)$ fluctuates erratically, the growth of 
$\langle a\rangle=\langle y\rangle xa_0\sim xa_0$ will be determined by the 
harmonic mean, denoted by $\mu(t)$. For this reason, a running harmonic mean of $\nu(t)$ was calculated, 
using a Gaussian weight with variance $\Delta t$:
\begin{equation}
\label{eq: 3.3}
\frac{1}{\mu(t)}=\frac{1}{\sqrt{2\pi}\Delta t}\int_0^{t_{\rm max}}{\rm d}t'\  \exp[-(t-t')^2/2\Delta t^2]\,\frac{1}{\nu(t')}
\ .
\end{equation}
The evolution of (\ref{eq: 3.1}) was followed for two different choices of initial distribution, namely
\begin{eqnarray}
\label{eq: 3.4}
p_0(a)=\frac{27}{2}a^2\exp(-3a)&\quad\quad\quad&{\tt PDF\ 1}
\\
\label{eq: 3.5}
p_0(a)=3Ca^2\exp\left(-Ca^{1/3}\right)&\quad\quad\quad&{\tt PDF\ 2}
\end{eqnarray}
with $C=[\Gamma(4/3)]^3$ in (\ref{eq: 3.5}). 
These distributions satisfy a requirement that $p_0(a)/a^2$ has a finite limit as $a\to 0$, 
which ensures that $\nu(t)$ is well-behaved at the start of the simulation.
Both distributions also satisfy $\langle a\rangle=1$.
For ${\tt PDF\ 1}$, equation (\ref{eq: 3.4}), the radius distribution has an exponential tail, 
and for ${\tt PDF\ 2}$, equation (\ref{eq: 3.5}), it is the volume distribution which has an exponential tail.

Figure \ref{fig: 1} shows plots of $\nu(t)$, over both short and long time intervals, 
for different distributions and choices of $\alpha$. While the initial evolution of $\nu(t)$ is smooth, at 
long times there are apparently chaotic fluctuations of $\nu(t)$ with increasing variance. 
These are more pronounced for the larger value of $\alpha$.  
The harmonic mean $\mu(t)$ is also plotted as a function of $t$ for the long-time plots, including the result 
of using a different random number seed for comparison.  
It is not clear whether it is asymptotic to the value of $27/4=6.75$ suggested by the Lifshitz-Slyozov theory \cite{Lif+61}.

The arguments in section \ref{sec: 2} suggest that the fluctuations of $\nu$ will be of order 
$\Omega=\alpha x/\sqrt{N}$. For the case where the PDF of the initial radii is given by (\ref{eq: 3.4}), 
the values of $x$ and $N$ at different times and the value of $\Omega$ are given in table \ref{tab: 1}.
Comparison with figure \ref{fig: 1} confirms that $\nu $ is subject to strong fluctuations when $\Omega$ is large.

\begin{table}[ht]
\caption{Tabulation of parameters for the examples in figure \ref{fig: 1}({\bf a})-({\bf f}).}
\centering
\begin{tabular}{c c c c c}
\hline\hline
 $t$ & $\alpha$ & $x$ & $N$ & $\Omega=\alpha x/\sqrt{N}$\\ [0.5ex]
\hline
\\ [ 1 ex]
$80$   & $25.0$   &  $4.05$ &  $2.92\times 10^5$ & $0.19$ \\ 
$800$ & $25.0$   &  $7.94$ &  $4.07\times 10^4$ & $0.98$  \\
$8000$   & $25.0$   & $16.3$ &  $4.78\times 10^3$ & $5.9$  \\ 
$80$   & $2.5$   &  $4.02$ &  $3.91\times 10^5$ & $0.016$ \\ 
$800$ & $2.5$   &  $7.87$ &  $5.67\times 10^4$ & $0.083$  \\
$8000$   & $2.5$   & $16.2$ &  $6.89\times 10^3$ & $0.49$  \\ 
\hline
\end{tabular}
\label{tab: 1}
\end{table}

The numerical integration of (\ref{eq: 3.1}) used a simple Euler scheme. The data in figure \ref{fig: 1} used 
timestep $\delta t=0.01$. Varying the timestep changed the numerical values, but not the qualitative character of the 
plots. The numerical evaluation of $\nu$ used an estimate $\dot x(t)=[x(t+\delta t')-x(t)/\delta t']$ with $\delta t'=0.1$.

\begin{figure}
\begin{center}
\includegraphics[width=6.3cm]{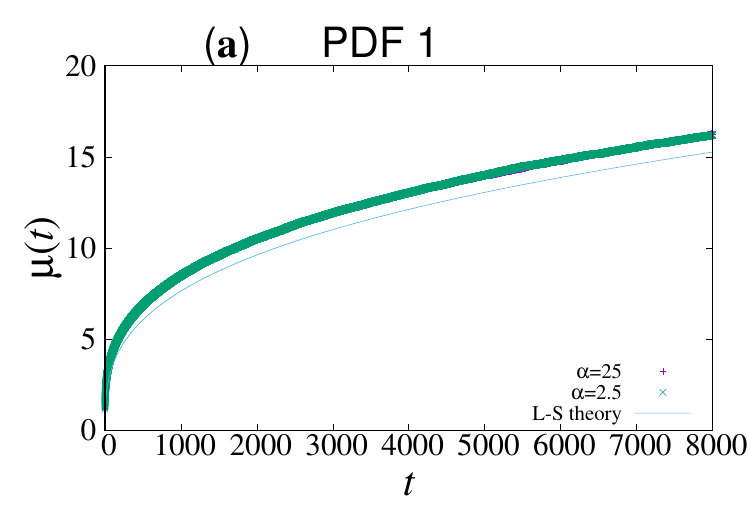}
\includegraphics[width=6.3cm]{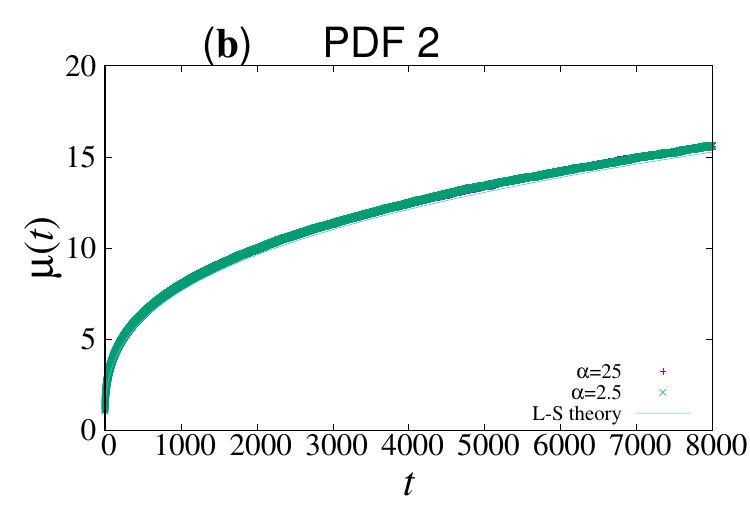}
\end{center}
\caption{
\label{fig: 2}
Growth of  $\langle a(t)\rangle$ compared with Lifshitz-Slyozov prediction. 
({\bf a}) Radius distribution has exponential tail (equation (\ref{eq: 3.4})).
({\bf b}) Volume distribution has exponential tail (equation (\ref{eq: 3.5})).
}
\end{figure}

Figure \ref{fig: 2} shows the evolution of the mean droplet radius, $\langle a\rangle$, compared 
with the prediction of the Lifshitz-Slyozov theory:
despite $\nu(t)$ having wild fluctuations, the growth of the mean radius is quite close to the Lifshitz-Slyozov prediction, 
and there is no significant difference between the results for $\alpha=2.5$ and $\alpha=25$. 
 
\begin{figure}
\begin{center}
\includegraphics[width=6.3cm]{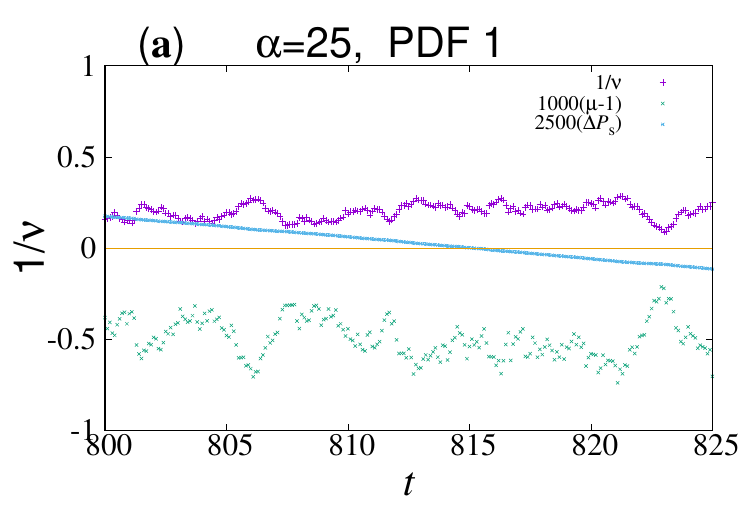}
\includegraphics[width=6.3cm]{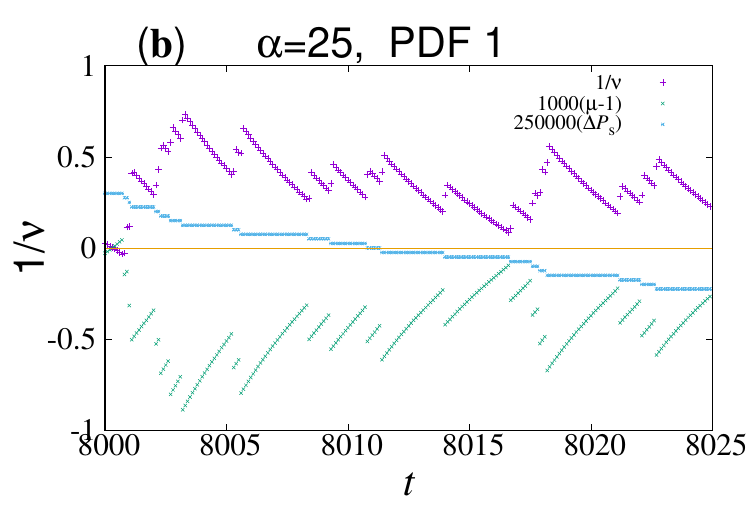}
\end{center}
\caption{
\label{fig: 3}
The same data sets as for figure \ref{fig: 1}({\bf e}), plotted on shorter intervals, showing fluctuations of $\nu(t)$,
comparing them with fluctuations of $\langle y\rangle-1$ and survival probability $P_{\rm s}(t)$ (with the local 
average over the interval subtracted). 
({\bf a}) Intermediate time: $t\in[800,825]$.
({\bf b}) Late stage: $t\in[8000,8025]$.
}
\end{figure}

\begin{figure}
\begin{center}
\includegraphics[width=6.3cm]{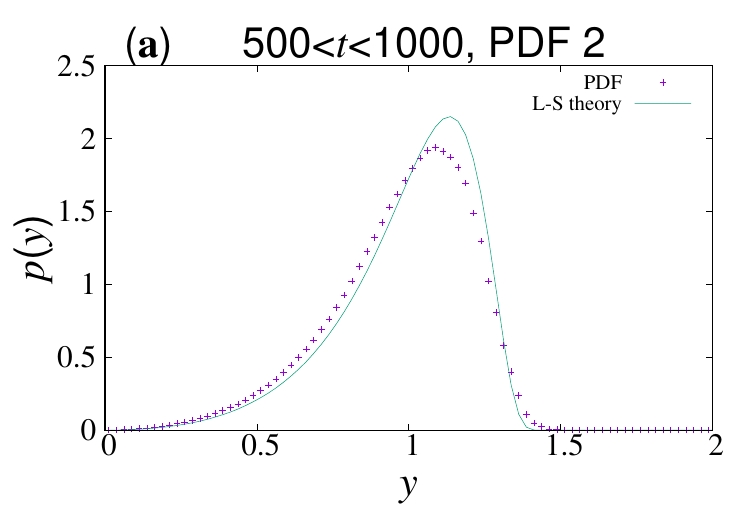}
\includegraphics[width=6.3cm]{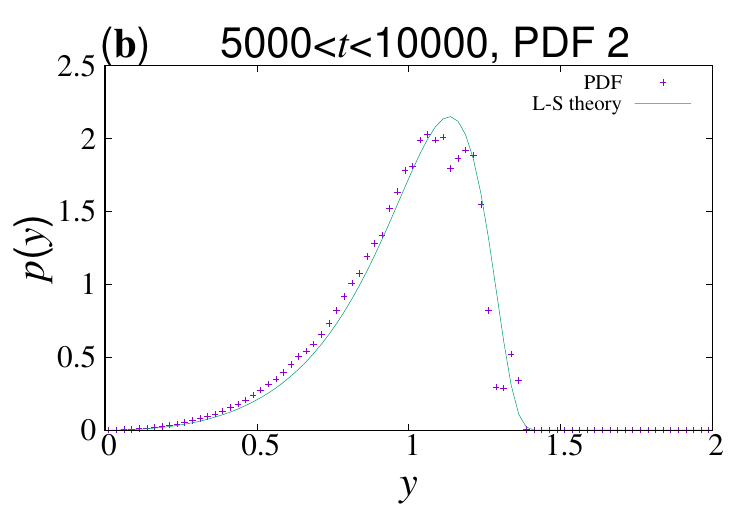}
\end{center}
\caption{
\label{fig: 4}
Distribution of the scaled droplet size, $p(y)$, using the same data sets as for figure \ref{fig: 1}({\bf e}). 
The PDFs are compared with the Lifshitz-Slyozov distribution, (\ref{eq: 3.6}). The 
PDF was accumulated for two different intervals: ({\bf a}), $t\in[500,1000]$, ({\bf b}), $t\in [5000,10000]$.
}
\end{figure}

Figure \ref{fig: 1} shows pronounced and increasing fluctuations of $\nu (t)$. The character of these 
fluctuations changes as time increases, as illustrated in figure \ref{fig: 3} (which displays $1/\nu(t)$ 
rather than $\nu(t)$, because the latter diverges when there are zeros of $\dot x$). At intermediate time, the fluctuations 
resemble Ornstein-Uhlenbeck noise ({\bf a}). At large times ({\bf b}) there is a sequence of abrupt increases corresponding 
to the evaporation of individual droplets as discussed in section \ref{sec: 2.3}, followed 
by decreases with approximately equal gradient. Figure \ref{fig: 3} also displays $1-\langle y\rangle$, which has fluctuations 
which mirror those of $\tilde \nu=1/\nu $, as expected from equation (\ref{eq: 2.19}). 
The deviation $\Delta P_{\rm s}$ of the survival probability from its average value in the interval is also plotted.
The latter shows that the abrupt increases of $1/\nu$ in figure \ref{fig: 3}({\bf b}) are 
a consequence of individual droplets evaporating. Both 
$\langle y\rangle-1$ and $\Delta P$ are multiplied by large factors to match the scale of the plot.

Figure \ref{fig: 4} shows the PDF of $y=as$, for the data in figure \ref{fig: 1}({\bf e}), accumulating data 
over different time intervals. These distributions are compared with the distribution predicted by the 
Lifshitz-Slyozov theory \cite{Lif+61}:
\begin{equation}
\label{eq: 3.6}
p(y)=
\left\{
\begin{array}{cc}
\frac{3^4}{2^{5/3}} 
\frac{y^2\exp\left[1-\frac{1}{1-2y/3}\right]}{(y+3)^{7/3}(\frac{3}{2}-y)^{11/3}}& y<\frac{3}{2}\cr
0 & y>\frac{3}{2}
\end{array}
\right.
\end{equation}
The empirical distributions are close to the Lifshitz-Slyozov prediction, but there are significant differences.

These investigations show that the droplet growth rate parameter $\nu(t)$ exhibits 
erratic fluctuations, which increase with time, and also increase with the dimensionless 
parameter $\alpha$ (see figure \ref{fig: 1}). There is evidence that these fluctuations 
are associated with counting fluctuations: in particular, in the later stages of the evolution, there is 
evidence that fluctuations of $\nu(t)$ are associated with the evaporation of individual droplets (see figure \ref{fig: 3}({\bf b})).
Despite the behaviour of $\nu (t)$ being very different from the Lifshitz-Slyozov prediction, the average 
droplet size is in quite good agreement with the Lifshitz-Slyozov prediction, as a consequence 
of the running harmonic mean of $\nu(t)$ approaching values which are close to $27/4$ (figure \ref{fig: 2}). 
The droplet size PDF is close to the Lifshitz-Slyozov prediction, but shows systematic differences (figure \ref{fig: 4}). 

\section{Simulation of reduced equations of motion}
\label{sec: 4}

It was argued in section \ref{sec: 2} that the dimensionless parameter 
$\alpha$ is typically very large, and that this implies that the scaled 
droplet size $y$ should satisfy $\langle y\rangle -1\ll1$. It is possible 
to impose $\langle y\rangle =1$ as a constraint, and dispense with 
the equation of motion for the supersaturation $\Phi_{\rm s}(t)$, or for the 
dimensionless critical droplet radius $x(t)$. This approach is equivalent 
to that of \cite{Pen97,Nie+99,Car+04}, but it might be expected that, if the initial droplet 
sizes are drawn at random, the effect of counting number fluctuations may be 
especially severe. This section will describe an approach which avoids 
treating the equation for the supersaturation, obtaining an expression 
for the value of $\tilde \nu=1/\nu$ directly, equation (\ref{eq: 4.6}) below. 
It will be shown that integrating the resulting equations leads to very severe 
fluctuations if the initial droplet radii are random.
 
Consider the condition for the expectation value of the scaled droplet radius $y$
to remain equal to unity. The time-derivative of of the mean value (\ref{eq: 2.20}) is
\begin{equation}
\label{eq: 4.1}
\frac{{\rm d}\langle y\rangle}{{\rm d}\tau }=\langle v_y\rangle +\lambda\langle y\rangle
\end{equation}
where $\tau$ is the time variable introduced in (\ref{eq: 2.14}) and where
\begin{equation}
\label{eq: 4.2}
v_y=\frac{{\rm d}y}{{\rm d}\tau}
\ ,\ \ \ 
\lambda=-\frac{1}{N}\frac{{\rm d}N}{{\rm d}\tau}
\end{equation}
with $N$ treated as if it were a continuous variable. 
Now using equation (\ref{eq: 2.14}) in (\ref{eq: 4.2}), and recalling the constraint $\langle y\rangle=1$, 
leads to an explicit equation for $\tilde \nu$:
\begin{equation}
\label{eq: 4.3}
\tilde \nu=\lambda+\biggl\langle\frac{1}{y}\biggr\rangle-\biggl\langle\frac{1}{y^2}\biggr\rangle
\ .
\end{equation}
This equation is not very convenient as it stands because it involves 
expectation values of quantities which diverge as $y\to 0$. The expectation values 
are finite because the velocity also diverges as $y\to 0$, so that if $p(y,\tau)$ is the probability 
density of $y$, then $p(y,\tau)/y^2$ approaches a finite limit as $y\to 0$. A more convenient 
formulation is to use a variable proportional to the volume of the droplet
\begin{equation}
\label{eq: 4.4}
z=y^3
\end{equation}
so that the equation of motion for $z$ is
\begin{equation}
\label{eq: 4.5}
\frac{{\rm d}z}{{\rm d}\tau}=3\left[z^{1/3}-1-\tilde \nu z\right]
\end{equation}
(which is a variant of an expression in \cite{Lif+58,Lif+61}), and the corresponding equation for $\tilde \nu$ is
\begin{equation}
\label{eq: 4.6}
\tilde \nu=\langle z^{-1/3}\rangle-\langle z^{-2/3}\rangle+\lambda=0
\ .
\end{equation}

Equation (\ref{eq: 4.6}) provides a prediction for the 
growth rate parameter $\tilde \nu$ which does not require integration of the equation 
for supersaturation. It corresponds to the $\alpha \to \infty$ limit of the equations
which were integrated in section \ref{sec: 3}, which presented evidence that there are
fluctuations of $\nu(t)$ which become more pronounced as $\alpha$ increases. It might, therefore, be anticipated that 
integration of equation (\ref{eq: 4.6}) will exhibit a pronounced instability. 

\begin{figure}
\begin{center}
\includegraphics[width=10.0cm]{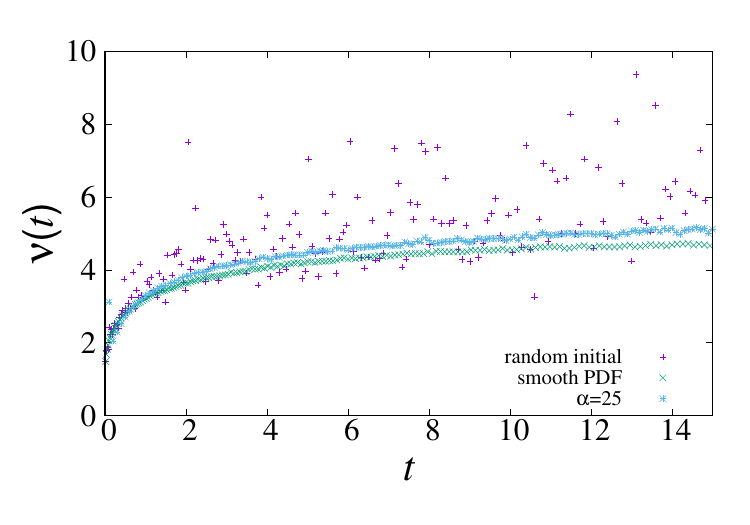}
\end{center}
\caption{
\label{fig: 8}
Evolution of $\nu(t)$ computed using equations (\ref{eq: 4.5}), (\ref{eq: 4.6}), compared
with reference data from figure \ref{fig: 1}({\bf e}) (initial PDF 2, $\alpha=25$, blue). 
The simulation of (\ref{eq: 4.5}), (\ref{eq: 4.6}) with the same random initial (purple) radii show a pronounced scatter almost 
immediately. For simulations with smooth initial distribution, $\nu$ depends smoothly 
upon $t$, initially following the reference case quite closely (green).
}
\end{figure}

Equation (\ref{eq: 4.5}) was integrated using equation (\ref{eq: 4.6}) to determine 
$\tilde \nu$ as a function of $\tau$. Equation (\ref{eq: 2.13}) was used to express
$t$ in terms of $\tau$. The value of $\lambda$ was estimated from the change in the survival 
probability over the timestep $\delta \tau$ of the numerical integration (at the short times considered in figure \ref{fig: 8}, 
the values of $N$ are sufficiently large that its discreteness has negligible effect). 
Figure \ref{fig: 8} compares $\nu(t)$ for different cases. 
The instabilities were so much greater that a much shorter time interval is displayed, up to $t=15$.
The data plotted in figure \ref{fig: 1}({\bf g}) (plotted in blue) are used as a reference (here $\alpha=25$, and 
the initial PDF is given by (\ref{eq: 3.5})): over this short interval, the erratic fluctuations 
seen in figure \ref{fig: 1}({\bf g}) are not yet developed. This is compared with data for the 
same random distribution of particle radii, but with $\nu(t)$ obtained by integration of (\ref{eq: 4.5})
and (\ref{eq: 4.6}) (purple). The latter data shows pronounced erratic fluctuations developing 
almost immediately.

As a control, figure \ref{fig: 8} also shows data for a case which suppresses counting fluctuations, by placing 
the initial droplet radii on a lattice, and assigning them a weight proportional to the
initial probability density. The case where the initial PDF corresponds to equation (\ref{eq: 3.5}) should be directly comparable with 
the data from figure \ref{fig: 1}({\bf g}). In this case the time dependence 
of $\nu(t)$ is smooth, (green curve in figure \ref{fig: 8}) and 
this simulation initially follows that reference quite closely, verifying equation (\ref{eq: 4.6}). 

The conclusion is that integration of (\ref{eq: 4.5}) and (\ref{eq: 4.6}) is extremely unstable if the initial 
droplet sizes are drawn from a random distribution.

\section{Concluding remarks}
\label{sec: 5}

This paper discussed the equations of motion for Ostwald ripening. The equation 
of motion for the supersaturation $\Phi_{\rm s}(t)$ (or equivalently, for the critical droplet size $a_0x(t)$), 
contains a dimensionless parameter (which was denoted by $\alpha$). This dimensionless parameter
is very large for the atmospheric aerosol, and probably for most potential contexts of Ostwald ripening.
In section \ref{sec: 2} it was argued that the equations of motion for the growth rate parameter $\tilde \nu=x^2\dot x$ 
may exhibit instability. One source is counting number fluctuations, which are amplified so that there are 
pronounced fluctuations of $\nu$ if $\Omega\equiv \alpha x/\sqrt{N}\gg 1$.

Numerical simulations reported in section \ref{sec: 3} show evidence for this instability. In the cases where the initial droplet 
distribution is drawn independently from a probability distribution, there are erratic fluctuations 
of  $\nu(t)$ which grow as both time and $\alpha$ are increased (figure \ref{fig: 1}). The running harmonic mean 
of $\nu(t)$ reaches values which are quite close to the Lifshitz-Slyozov prediction, and the mean droplet 
radius is close to the Lifshitz-Slyozov prediction (figure \ref{fig: 2}). At long-time scales the droplet size distribution
is close to, but significantly different from their prediction (figure \ref{fig: 4}). 

Many applications of Ostwald ripening will correspond to very large values of $\alpha$, and much of the 
literature (for example \cite{Pen97,Nie+99}) makes assumptions from the start which are equivalent to 
assuming the $\alpha\to \infty$ limit. In this limit, the equation of motion for $x(t)$ can be dispensed with, and 
replaced by an assumption that the mean value of the scaled droplet radius, $y=a/a_{\rm cr}$, satisfies
$\langle y\rangle =1$. This leads to an equation, (\ref{eq: 4.6}), for $\tilde \nu(t)$ which can be evaluated as the equation 
of motion for the scaled droplet volume, (\ref{eq: 4.5}), is evolved. However, if the droplet sizes are drawn from a random distribution, 
this approach produces wildly fluctuating values of $\nu(t)$ almost immediately, as illustrated in figure \ref{fig: 8}.

Taken together, these studies indicate that the theory of Ostwald ripening is incomplete, because the 
evolution of the growth rate parameter $\nu(t)$ is typically subject to erratic fluctuations, rather than approaching the constant
value $27/4$ as predicted in \cite{Lif+58,Lif+61}. This appears to have a significant effect upon the 
asymptotic droplet size distribution, $p(y)$. While it would be desirable to have a theory for the long-time limit 
of the distribution $p(y)$, these numerical studies do indicate that will be a difficult task, and that there may not 
be a unique asymptotic distribution.

\bmhead{Statements}

No externally sourced data was processed. The programs and data used to generate 
the figures are available from the author. No grants 
were received specifically for this work, and there are no relevant financial 
or non-financial interests to disclose.


\begin{thebibliography}{}

\bibitem{Ost96}
W. Ostwald, 
{\sl Lehrbuch der Allgemeinen Chemie}, vol. 2, part 1. Leipzig, (1896).

\bibitem{Lif+58}
I. M. Lifshitz and V. V. Slezov, 
{\it Zh. Eksp. Thoer. Fiz.}, {\bf 35}, 2, (1958).
English transl: {\sl Kinetics of decomposition of supersaturated solid solutions},
{\it Sov.Phys-JETP}, {\bf 8}, 331-39, (1959).

\bibitem{Lif+61}
I. M. Lifshitz and V. V. Slyozov,
{\sl The kinetics of precipitation from supersaturated solid solutions},
{\it J. Phys. Chem. Solids}, {\bf 19}, 35-50, (1961)

\bibitem{Wag61}
C. Wagner,
{\sl Theorie der alterung von niederschlagen durch umlosen (Ostwald-reifung)},
{\it Z. Elektrochem.}, {\bf 65}, 581-591, (1961)

\bibitem{Voo84}
P. W. Voorhees,
{\sl The theory of Ostwald ripening},
{\it J. Stat. Phys.}, {\bf 38}, 231-52, (1984).

\bibitem{Con26}
A. Conan Doyle,
{\sl The Adventure of the Blanched Soldier}, 
{\it Strand Magazine}, {\bf 72}, (1926).
(Similar formulations occur in other Sherlock Holmes stories).

\bibitem{Wil25}
M. Wilkinson,
{\sl Oscillatory Instability in an Ostwald Ripening Process},
arXiv 2503.18194, submitted to {\it J. Stat. Mech.: Theory Exp.}, (2025).

\bibitem{Che+93}
M. K. Chen and P. W. Voorhees,
{\sl The dynamics of transient Ostwald ripening},
{\it Modelling Simul. Mater. Sci. Eng.}, {\bf 1}, 591-612, (1993).

\bibitem{Yao+93}
J. H. Yao, K. H. Elder, H. Guo and M. Grant,
{\sl Thoeory and simulation of Ostwald ripening},
{\it Phys. Rev. B}, {\bf 47} 14110-14125, (1993).

\bibitem{Car+04}
J. A. Carrillo and T. Goudon,
{\sl A Numerical Study on Large-Time Asymptotics of the Lifshitz–Slyozov System},
{\it J. Sci. Comp.}, {\bf 20}, 69-113, (2004).

\bibitem{Pen97}
O. Penrose,
{\sl The Becker-D\"oring equations at large times and their connection 
with the LSW theory of coarsening},
{\it J. Stat. Phys.}, {\bf 89}, 305-320, (1997).

\bibitem{Nie+99}
B. Niethammer and R. L. Pego,
{\sl Non-Self-Similar Behavior in the LSW Theory of Ostwald Ripening},
{\it J. Stat. Phys.}, {\bf 95}, 867-902, (1999).

\bibitem{Rat+85}
L. Ratke and W. K. Thieringer,
{\sl The influence of particle motion on Ostwald ripening in liquids},
{\it Acta. metall.}, {\bf 33}, 1793-1802, (1985).

\bibitem{Bal02}
A. Baldan,
{\sl Progress in Ostwald ripening theories and their applications to nickel-base superalloys
Part I: Ostwald ripening theories},
{\it J. Materials Sci.}, {\bf 37}, 2171– 2202, (2002).

\bibitem{Bro89}
 L. C. Brown, 
 {\sl A new examination of classical coarsening theory},
{\it Acta Metallurgica}, {\bf 37}, 71-77 (1989).

\bibitem{Hil+92}
M.. Hillert, O. Hunderi and  N. Ryum,
{\sl Instability of distribution functions in particle coarsening},
{\it Scripta Metallurgica}, {\bf 26}, 1933-1938, (1992).

\bibitem{Mee+96}
B. Meerson and P. V. Sasorov,
{\sl Domain stability, competition, growth, and selection in globally constrained bistable systems}
{\it Phys. Rev. E}, {\bf 53}, 3491-3494, (1996).

\bibitem{Gir+98}
B. Giron, B. Meerson and P. V. Sasorov
{\sl Weak selection and stability of localized distributions in Ostwald ripening},
{\it Phys. Rev. E}, {\bf 58}, 4213-4216, (1998).

\bibitem{Mee99}
B. Meerson,
{\sl Fluctuations provide strong selection in Ostwald ripening},
{\it Phys. Rev. E}, {\bf 60}, 3072-3075, (1999)

\bibitem{Nie+01}
B. Niethammer and R. L. Pego,
{\sl The LSW Model for Domain Coarsening: Asymptotic Behavior for Conserved Total Mass},
{\it J. Stat. Phys.}, {\bf 104},  1113-1144, (2001).

\bibitem{Nie+06}
B. Niethammer and J. J. L. Vel\'azquez,
{\sl On the Convergence to the Smooth Self-similar Solution in the LSW Model},
{it Indiana Uni. Math. J.}, {\bf 55},  761-794, (2006).

\bibitem{Col+02}
J-F. Collet, T. Goudon and A. Vasseur,
{\sl Some Remarks on Large-Time Asymptotic of the Lifshitz–Slyozov Equations},
{\it J. Stat. Phys.}, {\bf 108}, 341-359, (2002).

\bibitem{Mas71}
B. J. Mason,
{\sl The Physics of Clouds}, 2nd. ed.,
Oxford, University Press, (1971).

\bibitem{Mar+84}
J. A. Marqusee and  J. Ross,
{\sl Theory of Ostwald ripening: Competitive growth and its dependence on volume fraction},
{\it  J. Chem. Phys.}, {\bf 80}, 536-543, (1984).

\bibitem{Gre56}
G. W. Greenwood,  
{\sl The growth of dispersed precipitates in solutions},
{\it Acta. Met.}, {\bf 4}, 243-48, (1956).

\end{thebibliography}
\end{document}